\documentclass[%
 reprint,
 amsmath,amssymb,
 aps,
]{revtex4-2}

\usepackage{graphicx}
\usepackage{dcolumn}
\usepackage{bm}
\usepackage{hyperref}

\begin{document}

\title{Anomalous mass dependency in Hydra endoderm cell cluster diffusion}
\author{Aline Lütz, Carine Beatrici, Lívio Amaral and Leonardo Brunnet\footnote{E-mail: leon@if.ufrgs.br}}
\affiliation{Instituto de Física, Universidade Federal do Rio Grande do Sul, Av. Bento Gonçalves 9500, Porto Alegre, RS, Brazil.}
\date{\today}

%

\begin{abstract}
Tissue organization plays a crucial role in morphogenesis, wound healing and cancer metastasis. Hydra, known for its regenerative capabilities, serves as an excellent model for studying cellular structures, particularly in pattern formation and the diffusion of cell clusters. Established experimental protocols enable the examination of regeneration and tissue segregation through optical microscopy.
Recent experiments challenge previous theories by suggesting that cell collective behavior is a key factor in hydra cell aggregate organization. Utilizing image treatment techniques on fluorescent images from Hydra's segregation experiment, we track endoderm cell clusters and investigate the relationship between their diffusion constant and cluster size. In contrast to results observed in non-active matter, we discover a nearly constant dependence of diffusion on cluster mass. This finding contributes to understanding the rapid segregation observed in recent experiments and holds implications for general cellular tissue processes.
\end{abstract}
\maketitle

\section{Introduction}
The mechanisms responsible for how cells migrate and organize themselves in multicellular biological environments involve a series of phenomena, only a part of which is understood. These mechanisms play a central role in tissue structuring during the morphogenesis process, wound healing, and even the proliferation of tumor cells throughout the body in the process of metastasis. This system operates at multiple scales, with feedback loops: the forces between cells trigger genetic mechanisms, which alter macroscopic parameters such as viscosities and tensions, which, in turn, further affect cellular-scale dynamics \cite{HANNEZO201912}.

\begin{figure*}
    \centering  
    \includegraphics[width=0.8\linewidth]{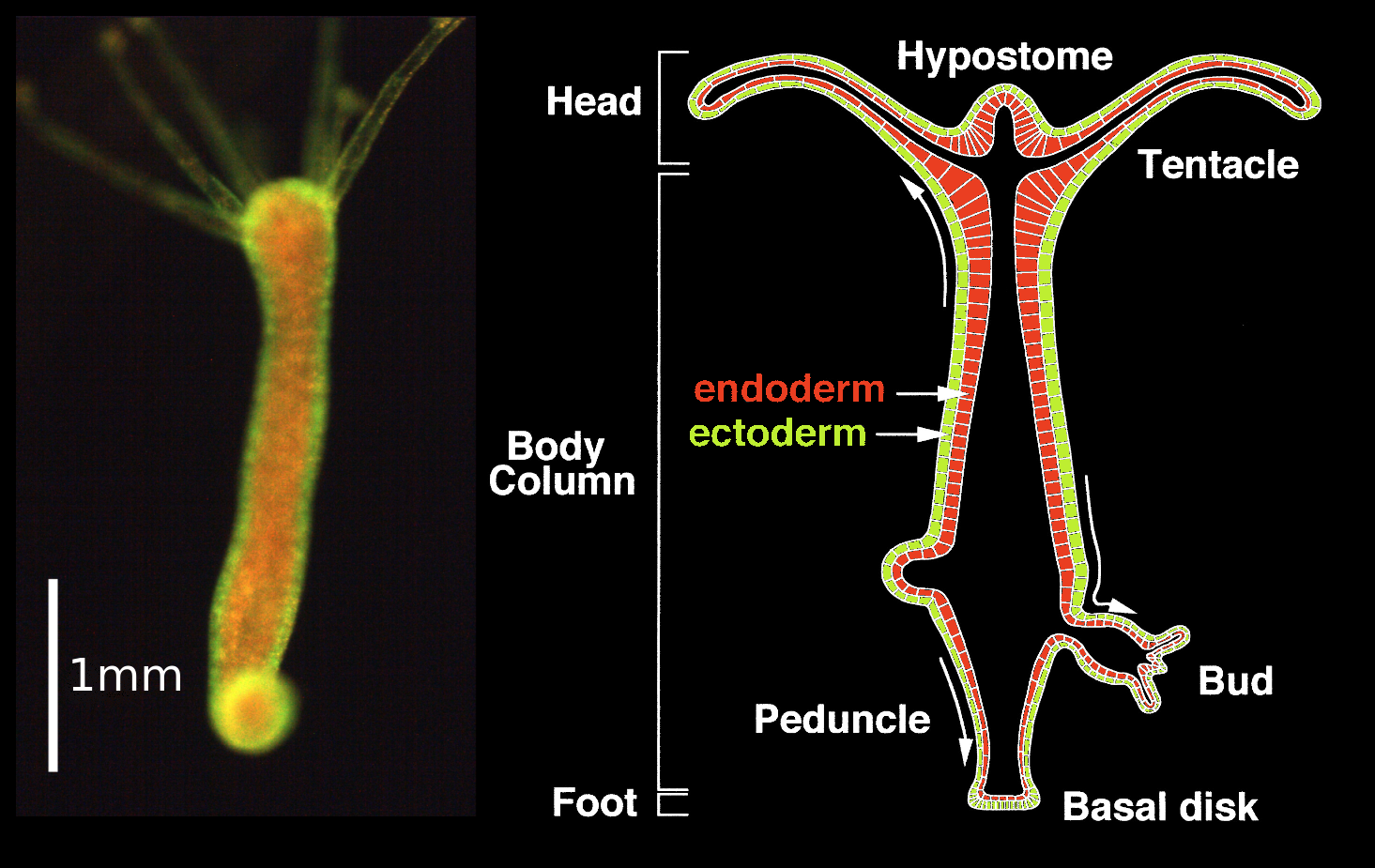}
    \caption{(left) \textit{Hydra vulgaris} genetically modified to be fluorescent. Hydras are simple animals, composed of
    two main tissues. The internal cells produce a red
    glow when exposed to green (or blue) light. The external cells produce a green glow when exposed to 
    blue light. (right) Schema detailing the morphology of a Hydra.}
    \label{fig:hydras}
\end{figure*}

Certain animals are known for their regenerative abilities, capable of reconstructing themselves even after their cells are completely separated. This capacity, combined with their simple structure, makes hydras an appropriate model for the study of cellular structures, particularly in pattern formation and diffusion of cell aggregates \cite{cerame-vivas1961}. Hydras, which are freshwater cnidarians, possess solely two types of embryonic tissues: endoderm (internal tissue) and ectoderm (external tissue). The left side of Fig. \ref{fig:hydras} presents a fluorescent microscope image of \textit{Hydra vulgaris}, while the right side provides a detailed scheme of its morphology.
 In the 1970s, in vitro protocols were developed \cite{Gie1972} to allow experiments on the initial stages of regeneration, the segregation of tissues \cite{TAKA2005}. This opened up the possibility of measuring the growth of cell aggregates via optical microscopy.

The initial theoretical perspective for describing tissue separation, conceived by Steinberg \cite{steinberg}, compared this phenomenon to oil coalescence in water, attributing separation essentially to the effects of aggregate (drop) fusion, minimizing surface tension, followed by diffusion, new fusion, and so on. However, experimental work from the early last decade \cite{mehes2012} indicated that the temporal evolution of the characteristic size of cell aggregates scaled with exponents higher than expected for oil coalescence in water. Similarly, recent experimental work by O. Escartin and collaborators \cite{escartin2017} suggests that the tension between different tissue types would be sufficient to explain the organization patterns in hydra cell aggregates, as proposed by Steinberg, but insufficient to explain the aggregate growth rates.

Steinberg's description implies the idea of thermal diffusion of aggregates. For oil and water, thermal fluctuations would be the driving force of diffusion, while for cell aggregates, it would be membrane fluctuations \cite{mom1995}. In the former case, aggregate diffusion clearly scales with the inverse of its mass, due to the absence of correlations between these fluctuations \cite{hua1963}, but in the latter, this is less clear. In fact, this view changes with the introduction of the concept of self-propelled particles by Vicsek \cite{vicsek1995}. He showed that by introducing an alignment tendency among particles, the system transitions from a gaseous phase to a phase of collective motion, with large groups moving coordinately.

The concept has found success in describing the dynamics of bird flocks, fish schools, and cellular clusters \cite{szabo2006}. The collective dynamics observed in the diffusion of aggregates within active matter simulations \cite{leon2008,beatrici2011} hint at the potential applicability of this concept to cell aggregates. It raises the perspective that cellular processes linked to embryogenesis, cancer, and wound healing could benefit from  collective motion, as suggested by simulations and experiments \cite{mones2015}. 

Investigating the diffusion of clusters and its size-related behavior could provide valuable insights into the collective motion contribution. However, there's currently a lack of direct experimental evaluation regarding aggregate diffusion and its relationship with size. Thus, the main goal in this work  is to explore this scenario by analyzing the diffusion of cell aggregates in Hydra segregation experiments.

Section 2 will involve detailing the hydra culture and sample preparation protocols, outlining the experimental methods for image capture, and presenting the computational tools used for image processing.
Section 3 will show the experimental findings and their analyses. Section 4  draws conclusions. 
Appendices \ref{culture}, \ref{processing} and \ref{analysis} furnish additional information on hydra culture, experimental setup, and details regarding image acquisition and processing.

\section{Method}
In our investigation into the diffusion properties of hydra's cellular clusters, we conducted a sequence of six experiments. Each experiment yielded a collection of images portraying migrating endodermic cell aggregates, surrounded by ectodermic cells, across a specific time interval. Our pipeline involved distinct stages to extract essential data from these experimental images. This section delineates our methodology for each step: conducting experiments, processing images, performing image analysis, and extracting measurements.

\begin{figure*}
    \centering
    \includegraphics[width=0.9\textwidth]{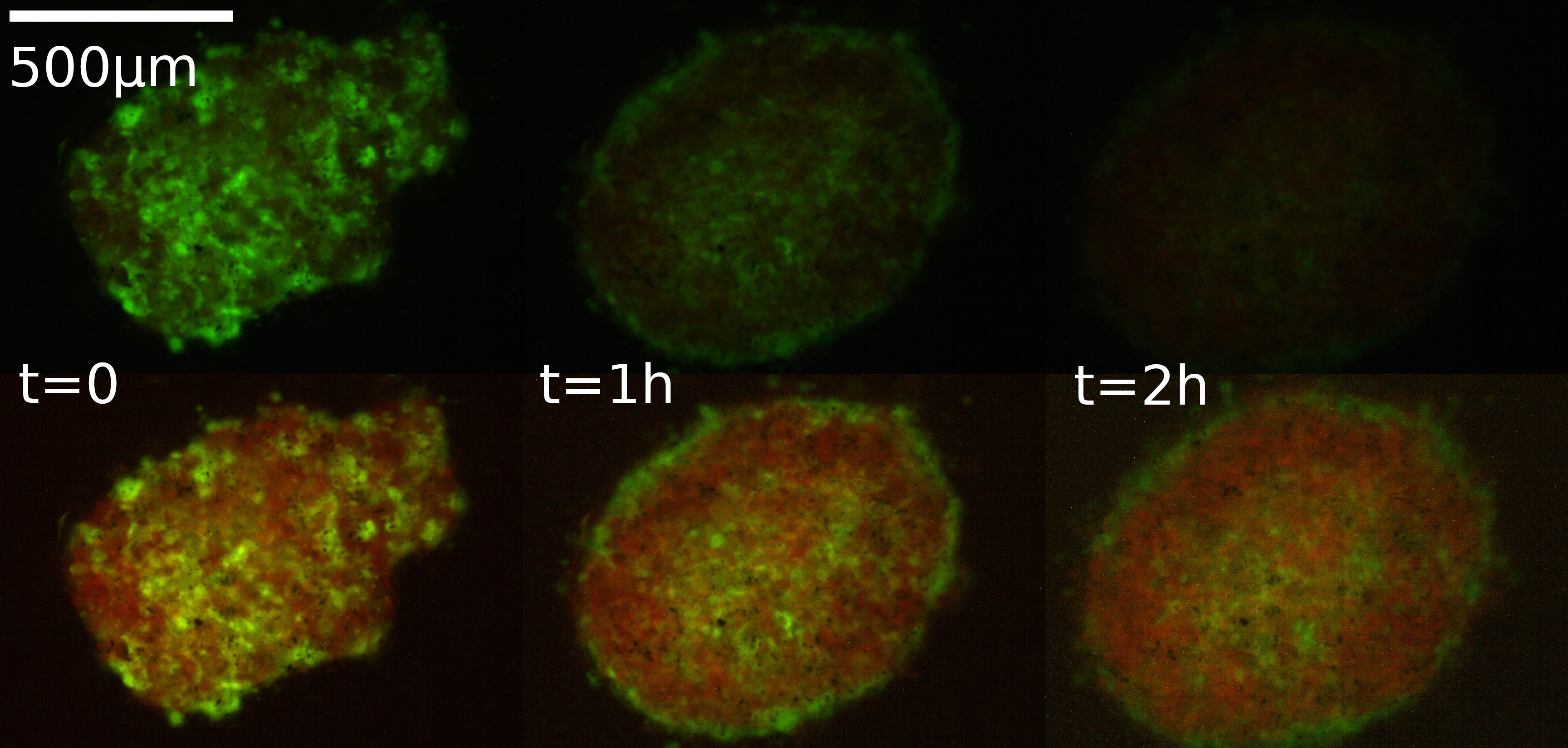}
    \caption{Snapshots during segregation process. Bleaching compensation example, the fluorophores loose glow as they are exposed to
    light, to compensate this effect we process the images in order to keep the average brightness 
    constant. At this point, red channel brightness was already increased in all images to 
    get to the same level as the green fluorescence gets.}
    \label{fig:bleaching-compensation}
\end{figure*}

The experimental setup involved hydra specimens of the \textit{vulgaris} species genetically modified to fluoresce in green (GFP) and red (RFP). The endodermal cells emitted red fluorescence under green light exposure, while ectodermal cells emitted green fluorescence under blue light exposure, as shown in Fig.\ref{fig:hydras}. The experimental protocol, based on the work of Lommis and Lenhof \cite{lenhoff}, is detailed in Appendix \ref{experiment}.

For regeneration, all cells from the hydra's body column were separated, mixed, centrifuged, and incubated using hydra medium (HM) and a dissociation medium (DM). Preceding the experiments, hydra samples underwent a fasting period of 24 to 48 hours to ensure an empty digestive cavity during the experimental cutting process\cite{kishimoto}.

During experiments, we captured time-series images, as in Fig. \ref{fig:bleaching-compensation}, to monitor cluster behavior and measure diffusive properties. However, raw experimental images were laden with extensive information—featuring both internal and external cells—alongside substantial noise. So, we processed the images to compensate bleaching, eliminate noise and retain pertinent information. The resultant images were binary representations, denoting the presence (value one) or absence (value zero) of endodermic cells. Please, see Fig. \ref{fig:image-processing} in appendix \ref{processing}.

Further analysis include identification and monitoring of the evolution of cell clusters within the processed images. This involved mapping the measured cluster area to the number of cells (proportional to mass, up to a constant, henceforth referred to as mass) and aligning the geometrical center with the center of mass for each cluster (refer to subsection Single Image Measurements in Appendix \ref{single-image} for details).

Ultimately, we quantified the diffusive properties of these clusters, including their mean squared displacement.
We analyze the characteristics of these cell clusters using KMeans, an unsupervised machine learning data grouping technique. For a detailed explanation, kindly refer to Appendix \ref{analysis}.

\subsection{Mean Squared Displacement (MSD)}
Once the trajectory of the center of mass of an aggregate has been defined, a windowing method is used to determine the evolution of its root mean square deviation. The idea behind this technique is to consider each point on the trajectory $r(t)$ as a possible starting point for the measurement and calculate the quadratic deviation with respect to this position one time $\Delta t$ later, i.e. $(r(t+\Delta t)-r(t))^2$. The procedure is repeated for the next interval $\Delta t$, $(r(t+2\Delta t)-r(t+\Delta t))^2$ and so on. Thus, if the total diffusion time of this aggregate is $T$, we will have $n\equiv T/\Delta t$ measures of the quadratic deviation for this time interval. We  calculate then the MSD for the interval $\Delta t$ as the mean defined by:
$$\Delta r^2(\Delta t) = \frac{1}{n} \sum_{i=0}^{n-1} (r((i+1)\Delta t)-r(i\Delta t)^2 \;. $$
In this method, the statistical sampling for calculating the MSD is large 
when $\Delta t << T$. As the interval $\Delta t$ increases, this sampling decreases. Therefore, more fluctuations in the MSD are expected for values of $\Delta t$ of the order of the time length of the trajectory, $T$.  

\section{Results}
\begin{figure*}
    \centering
    \includegraphics[width=\textwidth]{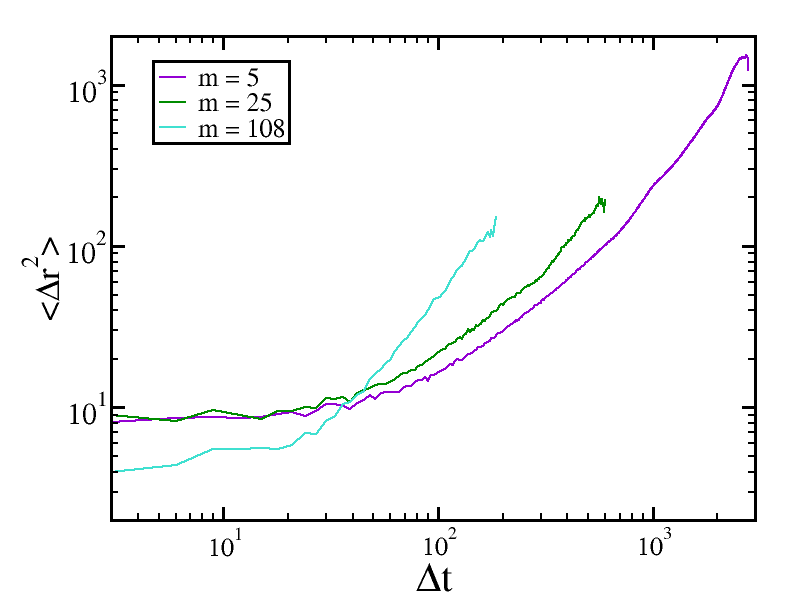}
    \caption{Curves of Mean Squared Displacement (MSD) (in $\mu m^2$) versus time lag ($\Delta t$) (in seconds) on a logarithmic scale on both axes. Different curves represent aggregates of different masses. Note a nearly horizontal trend for $\Delta t <$ 30 s. The fluctuations at the end of the curves for long $\Delta t$ are due to the windowing method used.}
    \label{fig:msd}
\end{figure*}

Figure \ref{fig:msd} depicts the evolution of the Mean Squared Displacement (MSD) for the center of mass of three aggregates with varying masses. In the early stage ($\Delta t < 30s$), a nearly horizontal trend is noticeable, primarily due to noise. With the passage of time, a distinct diffusive pattern emerges, and over extended durations, fluctuations rise, consistent with expectations based on the employed windowing method for MSD calculation.

Next, disregarding the horizontal portion of the curves, we fit the mean squared displacement using power law equations of the form:
\begin{equation}
\Delta r^2 = D \Delta t^\alpha
\label{power-law-fit}
\end{equation}
where the exponent $\alpha$ indicates whether the process is diffusive ($\alpha=1$), subdiffusive ($\alpha<1$), or superdiffusive ($\alpha>1$), and $D$ is the associated diffusion coefficient.
\begin{figure*}
    \centering
    \includegraphics[width=\textwidth]{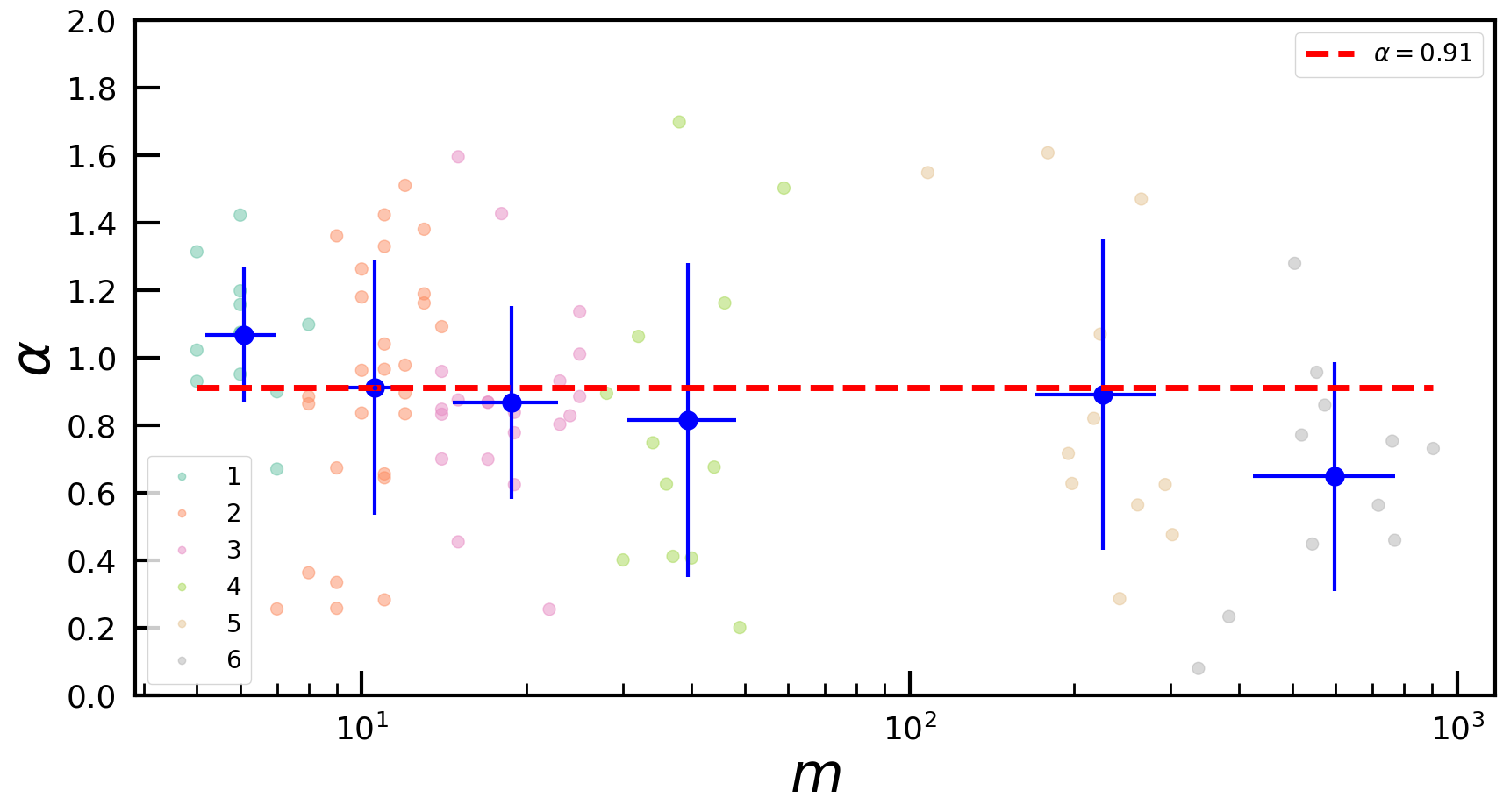}
    \caption{Exponent (\protect{$\alpha$}) fitted for the MSD as a function of the mass in the aggregates. The points in blue with error bars correspond to averages over mass groups as indicated in table \ref{tab:cluster}. The red dashed line represents the mean value of the exponent $\alpha$, with the last set of points excluded. }
    \label{fig:mass-alpha}
\end{figure*}

\begin{figure*}
    \centering
    \includegraphics[width=\textwidth]{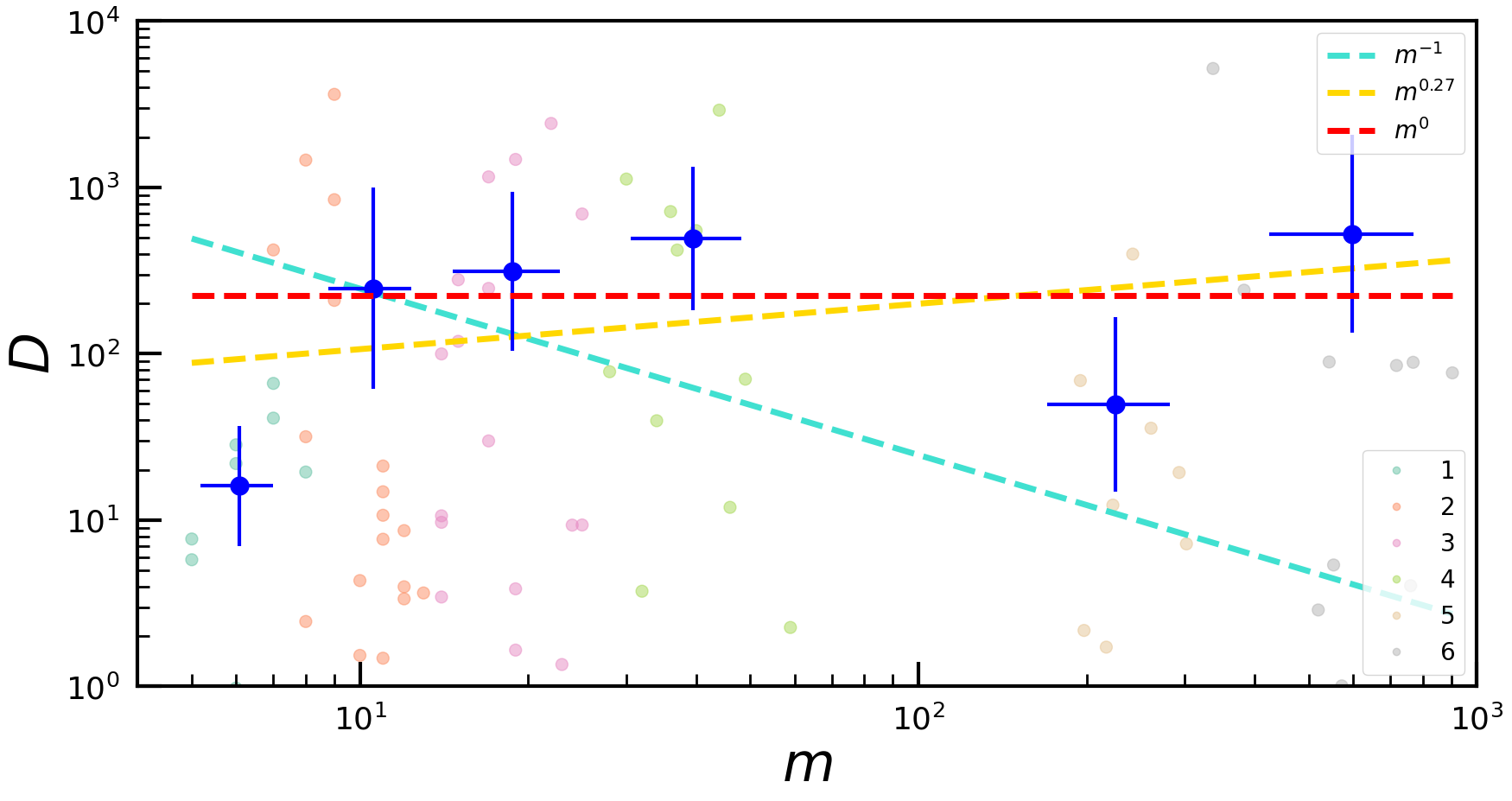}
    \caption{Diffusion constant (in $\mu m^2/s$) fitted as a function of aggregate mass. As in the previous figure, the points shown correspond to averages over the groups indicated in table \ref{tab:cluster}. The red dashed line is a linear fit $(D \sim m^{0})$ of the group's average (in blue), and the cyan dashed line represents the case where $D \sim m^{-1}$, assuming equal diffusion for an aggregate of the group id=2 of table \ref{tab:cluster}. } 
    \label{fig:mass-d}
\end{figure*}

By applying equation \ref{power-law-fit} to the dataset, we obtain the diffusion constant, $D$ and the exponent $\alpha$ for each cell aggregate. Consequently, each aggregate is defined by its mass ($m$), diffusion constant ($D$), and diffusion exponent ($\alpha$). Due to the limited number of aggregate data points (a total of $94$), which is inadequate for establishing a continuous mass function, we choose to partition the data into groups with similar characteristics. Subsequently, we analyze diffusion properties within these groups.
Employing the KMeans grouping algorithm~\cite{kmeans1957}, an unsupervised machine learning technique implemented in Scikit-Learn~\cite{scikit-learn}, we utilize mass and exponent parameters for data grouping. We determine the maximum number of aggregates that predominantly sorts  based on mass rather than exponent, resulting in six data groups. The points are visualized in Figures \ref{fig:mass-alpha} and \ref{fig:mass-d}, with the grouping outcome indicated by point colors, each color denoting a distinct data group. When ordered by mass, the KMeans groups are as follows:
\begin{table}
  \centering
  \begin{tabular}{|c|c|c|c|}
    \hline 
    Group id & Min Mass & Max Mass  \\
    \hline
    1 & 5 & 8 \\
    2 & 7 & 14 \\
    3 & 14 & 25 \\
    4 & 28 & 59 \\
    5 & 108 & 302 \\
    6 & 337 & 904 \\
    \hline
  \end{tabular}
  \caption{Machine learning data grouping result. The classification data includes mass and the diffusion exponent.
  The clusters are shown  Figures \ref{fig:mass-alpha} and \ref{fig:mass-d}. }
  \label{tab:cluster}
\end{table}
Figure \ref{fig:mass-alpha} displays the values obtained for the exponent $\alpha$, the points in blue are the averages obtained for each group.
The red dashed line points to  a slightly subdiffusive regime ($\alpha \approx 0.9 \pm 0.1$).

We now illustrate the relationship between the diffusion constant and aggregate mass. It is important to note that while $D$ remains constant in time  for a specific cluster, its numerical value may vary based on the aggregate mass. Traditional passive particle diffusion constants exhibit an inverse relationship with mass, expressed as $D(m) \sim m^{-1}$.

For certain active-collective particle aggregates, alternative diffusion-mass relations may emerge. Simulation results~\cite{carine2017,emanuel2021} indicate instances where diffusion shows no dependence on aggregate mass, denoted as $D(m) \sim m^{0}$ for collective active aggregates. In Figure \ref{fig:mass-d}, the blue data points represent the average group value for $D$.

A power law fitting is illustrated by the yellow dashed line, expressed as $D(m) = 1.75 m^{0.27}$. In contrast, the red dashed line shows a fit to a  constant, where $D(m) = m^{0} = 224.05$. To provide further context, the cyan dashed line illustrates a scenario in which the diffusion coefficient decreases inversely with mass, assuming a diffusion constant matching that of group id=2.

\section{Conclusion}
In this work, we present results from experiments on the segregation of ectoderm and endoderm cells in Hydra vulgaris. Based on the outcomes of  experiments, we applied image analysis methods to distinguish endoderm aggregates from ectoderm aggregates. We monitored the temporal evolution of the center of mass of endoderm aggregates, categorizing them by their number of cells.

The progression of the Mean Squared Displacement (MSD) for endoderm aggregates, as illustrated in Fig. \ref{fig:msd}, exhibits two distinct phases. In the initial phase, which spans short durations (on the order of seconds), the MSD remains relatively constant, attributed to experimental errors. The subsequent phase, lasting from minutes to one hour, it follows a power law and forms the core focus of this study. This phase enables an exploration of the correlation between aggregate diffusion and their mass, evaluated by the aggregate size in terms of the number of cells. Additionally, there exists a third time scale beyond one hour or for aggregates exceeding approximately 1000 cells, where rounding processes become dominant \cite{Rieu2002}. These time and/or size limits were utilized as upper bounds in our data analysis.

Due to significant fluctuations in the fitting parameters, we use a clustering algorithm to divide the aggregates in groups, with the number of cells in clusters falling within the intervals [5:8], [7:14], [14:25], [28:59], [108:302] and [337:904], respectively. We calculated the parameter averages within these intervals.

Excluded the short-time interval, the MSD exhibits power-law behavior with exponents slightly below 1. In other words, within experimental error, both small and large aggregates exhibit normal diffusive behavior.

Conversely, the key outcome emphasized in this study is the absence of a decrease in the diffusion constant with aggregate size across different groupings. This behavior is contrary to the anticipated pattern observed in the diffusion of non-active particles, where a typical exponent of -1 characterizes the relationship between diffusion and mass. 

In the experiments, the separation process imposed leads to the loss of ectoderm cells, resulting in an excess of endoderm cells in numerous instances. Nevertheless, isolated aggregates of various sizes were identified diffusing in all cases, enabling us to assess the dependence of their diffusion on mass. The observed behavior lends support to the hypothesis that the alignment interaction\cite{carine2017, emanuel2021} modifies the dependence of the diffusion constant on aggregate mass, yielding values higher than expected for non-active particles.

\section{Acknowledgments}
The hydras used were generously provided by Olivier Cochet-Escartin from the Lumière Matière Institute at the Claude Bernard Lyon 1 University.

\appendix
  
  \begin{center}
    {\LARGE\bf{Appendix}}
\end{center}
Section \ref{culture} below delineates steps involved in the maintenance, preparation of hydra specimens for experimentation and the  process used during typical experiments. Sections \ref{processing} and \ref{analysis} are dedicated to image processing and analysis, respectively.

\section{Hydra Culture}
\label{culture}
We follow here the procedure introduced by Lenhof\cite{lenhoff}.
Hydras are kept at $18^{\circ}C$ in glass containers containing Hydra Medium (HM). This medium consists of a concentrated solution of $NaHCO_3$ (18.92\%), $MgCl2.6H2O$ (9.2\%), $MgSO_4$ (4.33\%), $KNO3$ (1.367\%), and ${CaCl2}.2{H2O}$ (66.22\%). For use, the solution is diluted at a ratio of 20 ml per 20 L of distilled water, with a pH adjusted to 7 (neutral). The hydras are fed once or twice a week with freshly hatched brine shrimp.

\subsection{Experimental Procedure}
\label{experiment}
For the procedure of a typical experiment we follow Kishimoto and coworkers\cite{kishimoto}:
\begin{enumerate}
    \item Using a scalpel, the head and foot of each hydra are separated from the body column. Only the body columns are placed in a Petri dish.
    \item The body columns are transferred to a test tube containing 5 ml of HM at 4°C using a pipette.
    \item They are left to settle, and the procedure is repeated three times, washing with 5 ml of 4°C HM each time.
    \item After the third wash, the solution is replaced with 5 ml of 4°C DM.
    \item The body columns are incubated in the test tube at 4°C for 20 minutes in DM.
    \item The body columns are sliced with a scalpel on a Petri dish.
    \item The dissociation medium (DM) is changed by adding a new batch at 4°C.
    \item The medium should become cloudy as dissociation begins. Suction and expulsion of the liquid with the cells are repeated to break the cell bonds, taking care not to introduce air bubbles during the process.
    \item The liquid is passed through a 40-micron diameter filter and then centrifuged for 7 minutes.
    \item The resulting aggregate is removed from the bottom of the centrifuge tube, placed in a Petri dish, and sliced.
    \item Small aggregates are aspirated with a hypodermic needle (0.35 mm in diameter) and deposited in 1 mm-wide wells on a glass slide.
    \item The slide with the wells is covered with another glass slide of the same size.
    \item The two slides are sealed together with Teflon tape.
    \item To observe under the microscope, the slides are placed in a Petri dish and immersed in HM.
    \item Image capture is performed using a UI-1490SE-C camera from IDS Imaging\textsuperscript{\texttrademark} (resolution 3840x2748, color CMOS) attached to an inverted IX70 Olympus\textsuperscript{\texttrademark} microscope.

\end{enumerate}

\section{Image Processing}\label{processing}
Below, we describe how we processed the images to identify these aggregates.
For the image analysis, we developed various programs in Octave\cite{octave} and Python\cite{python}. These programs allow for the automatic recognition of cell clusters and their movement within the cell population. All programs used in this work can be accessed in the public GitLab repository of the \href{https://gitlab.com/labcel/hidras/experimentos}{Laboratory of Cell Structures}.

\subsection{Channel Splitting}\label{channel-splitting}

\begin{figure*}
    \centering
    \includegraphics[width=1.\textwidth]{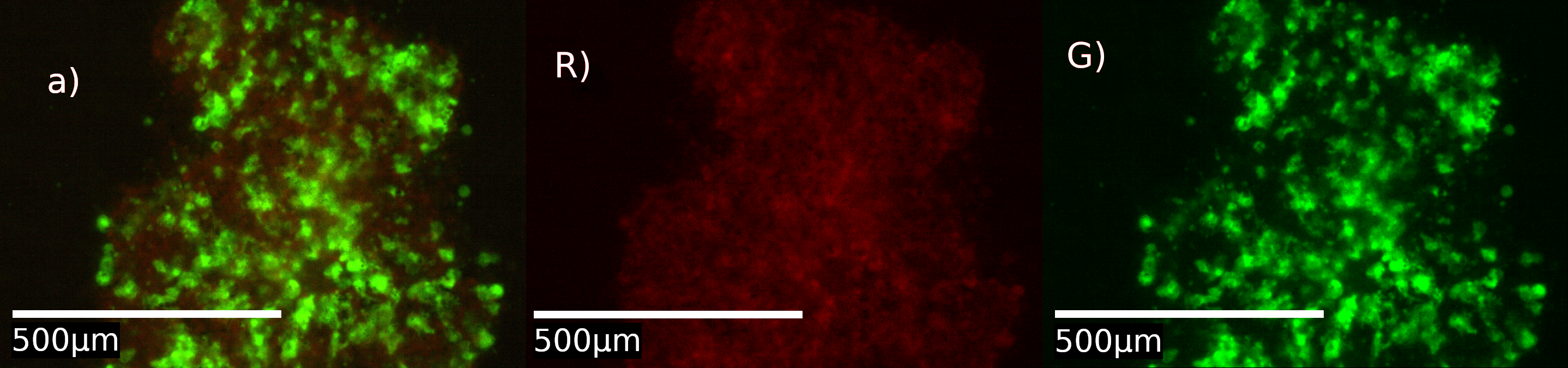}
    \caption{a) Original experimental image, external cells are marked in green while internal cells are 
    marked in red, even though the red fluorescence is much less bright than the green one. 
    R) Red Channel, only endodermic cells are visible in this image as they are the 
    only ones that glow in red. 
    G) Green channel, only ectodermic cells are visible in this image since they are the only ones
    that glow in green. 
    Blue channel images are not shown because they are all empty as neither cell types glow in this channel.
    We save each channel image data in gray scale, we keep the colors for aesthetic reasons.}
    \label{fig:split-rgb}
\end{figure*}

The images are in color, composed of red, green, and blue channels—commonly known as RGB. Each picture maintains a resolution of 3840x2748 points (or pixels). The intensity of each channel is measured with 16 bits, with values ranging between 0 and 255.
In figure 
\ref{fig:split-rgb}, we present an example of channel splitting, in  a) we have the original image, as taken by the microscope camera. The green cells (ectoderm) are directly visible, while the red ones (endoderm) glow much less. R) Red channel extracted from original image, the image is stored 
in gray scale, we present in red color for visual guide. Once the red channel is
isolated, the endodermic cells are more visible. We will process those images in order to improve them
before making analysis.
G) Green channel, as the green fluoresce domains the original image, the green channel image looks like
the original one.  In brief, red channel will have only internal cells  
while the green channel only external cells, 
at last, the blue channel (not shown) will not have any useful information.

As mentioned earlier, the red glow is weaker, so we strengthened the glow by boosting its intensity relative to the green color.

\subsection{Bleaching Compensation}
\label{bleaching}

The first step of image processing involves normalizing the brightness of the photos, 
as shown in Figure~\ref{fig:bleaching-compensation}.
The average brightness of each image is calculated and corrected to the reference value obtained at 
the beginning of the experiment. 
This adjustment is essential due to the significant brightness variation during
the experiments caused by fading fluorescence. 
The correction algorithm operates by normalizing all images based on the average 
intensity of the initial image.

\subsection{Binary Image}
\label{binary-image}

\begin{figure*}
    \centering
    \includegraphics[width=0.49\textwidth]{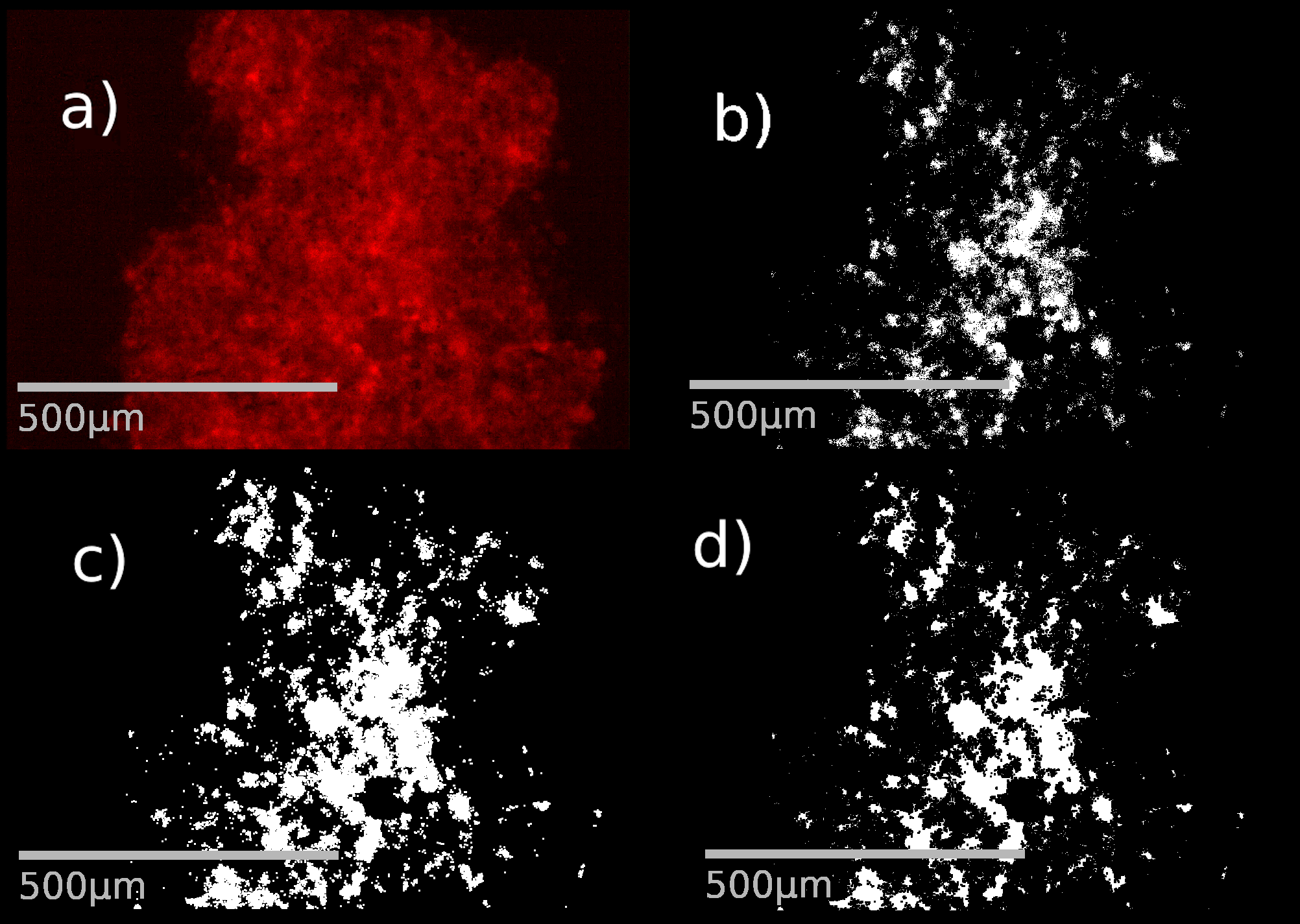}
    \caption{a) Red channel image processed to enhance and normalize the average brightness
    based on a standard image from the same experiment.
    b) Black and white version of the top-right image, the threshold is selected automatically.
    c) Image after applying the \textit{dilate} filter to the bottom-left image to smooth the aggregates. 
    d) Image after applying the \textit{erode} filter to the top-left image to reduce noise}
    \label{fig:image-processing}
\end{figure*}

To detect cell endoderm clusters, we utilize the red channel images and apply a transformation whereby each pixel is assigned a value of one if it belongs to an endodermic cell and zero otherwise, a process known as image binarization. An example of this binarization procedure is illustrated in Figure \ref{fig:image-processing}. The initial image comprises red channel intensities that have already been normalized for brightness. Using the Octave\cite{octave} \textit{graythresh} function, we determine the threshold that distinguishes between intensities denoting the absence or presence of a cell. Consequently, for each image, we obtain a binary matrix with a resolution of 3840x2748 pixels, wherein each element is either 0 or 1, as depicted in Figure \ref{fig:image-processing}b.

The binarization process can introduce discontinuities and roughness in the aggregate images. To address this, we utilize the Octave\cite{octave} \textit{dilate} filter, aiming to smooth edges and fill gaps, as depicted in Figure~\ref{fig:image-processing}c. In order to reduce noise—such as eliminating isolated points—we apply the Octave\cite{octave} \textit{erode} filter, illustrated in Figure~\ref{fig:image-processing}d. The sequence of \textit{dilate-erode} operations is iteratively performed on a standard experimental image until the resulting image matches the original.
The same number of repetitions is applied to subsequent images. 

\begin{figure*}
    \centering
    \includegraphics[width=0.8\textwidth]{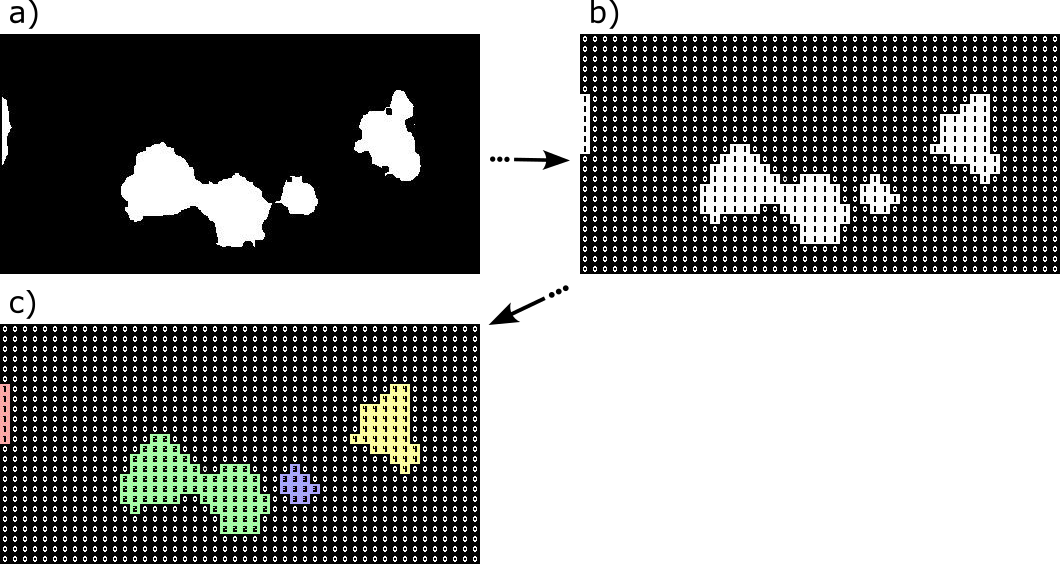}
    \caption{ a) Image converted to black and white, b) binary representation, c) each different aggregate  receives an index.}
    \label{fig:bw-labeling}
\end{figure*}

\section{Image Analysis}
\label{analysis}

Once all experimental snapshots have been processed, we move on to extracting the measurements of interest, including aggregate identification, center of mass, area, and displacement. As elaborated in the following two sections, these measurements can be categorized into two groups: those applicable to single images and those spanning multiple images. Initially, cluster identification is conducted independently for each image. However, to assess diffusive properties accurately, it's imperative to identify the same cluster across different snapshots.

\subsection{Single Image}
\label{single-image}
We employ the Octave\cite{octave} function \textit{bwlabel} to detect clusters within a single image. This function produces a matrix mirroring the input image's resolution, as shown in Figure \ref{fig:bw-labeling}c. Vacant spaces are represented as zeros, while each cluster is marked with a unique value. Relevant data, including the number of clusters, cluster area (in pixels), and the geometric center of each cluster, are directly extracted from these matrices.

The highest value within each matrix indicates the count of identified aggregates. The frequency of a specific value denotes the size of the corresponding aggregate. Each pixel's position in the matrix directly corresponds to its position in the image. Typically, a single cell's area approximates 1000 pixels, and the average cell diameter measures around 20 µm. Hence, each micrometer in our images corresponds to $1.8\mu m$. If required, the positional data of the geometric center of each aggregate in pixels can be converted to micrometers. To evaluate the number of cells within a cluster, we divide its area by the area of a single cell.

\subsection{Multiple Image}
\label{multiple-image}

The matrices derived from image processing offer pertinent data for monitoring clusters and tracking the evolution of their properties. Our Python-based\cite{python} tracking software, available on the \href{https://gitlab.com/labcel/hidras/experimentos}{Laboratory of Cell Structures} GitLab page, has been designed to identify aggregates in consecutive images. It analyzes similarities in both mass and the position of the center of mass between these aggregates. If these parameters fall within an appropriate range, their indices are then equated.
Subsequently, we can monitor and quantify both the displacement of clusters and variations in mass during occurrences of cluster fusion.

\bibliography{article}

\end{document}